\documentclass[twocolumn,secnumarabic,amssymb,amsmath, nofootinbib,tightenlines,nobibnotes,aps,prl,
superscriptaddress,10pt]{revtex4}
\usepackage{amsmath}%
\usepackage{longtable}%
\usepackage{bm}%
\usepackage{epsfig}

\newcommand{\tri}{| \! | \! |}
\newcommand{\rd}{{\rm d}}
\newcommand{\be}{\begin{equation}}
\newcommand{\ee}{\end{equation}}
\newcommand{\bey}{\begin{eqnarray}}
\newcommand{\eey}{\end{eqnarray}}

\newcommand{\bp}{{\bf p}}

\newcommand{\br}{{\bf r}}

\newcommand{\bRR}{{\bf R}}

\newcommand{\ph}{\varphi}

\renewcommand{\a}{\alpha}

\newcommand{\cU}{{\cal U}}

\newcommand{\bR}{{\mathbb R}}

\newcommand{\tr}{\mbox{Tr}}

\newcommand{\wt}{\widetilde}

\newcommand{\cE}{{\cal E}}

\input epsf

\newcommand{\donothing}[1]{}

\begin{document}

\title{Rigorous Derivation of the Gross-Pitaevskii Equation}
\author{L\'aszl\'o Erd\H os}
\affiliation{Institute of Mathematics, University of Munich,
Theresienstr. 39, D-80333 Munich, Germany}
\author{Benjamin Schlein}
\affiliation{Department of Mathematics, UC Davis, One Shields Ave,
Davis, CA 95616}
\author{Horng-Tzer Yau}
 \affiliation{Department of Mathematics, Harvard University,
One Oxford Street, MA 02138}
\date{7 December 2006}

\begin{abstract} The time dependent
Gross-Pitaevskii equation  describes the  dynamics of initially
trapped Bose-Einstein condensates. We present a rigorous proof of
this fact starting from a many-body bosonic Schr\"odinger equation
with a short scale repulsive interaction in the dilute limit. Our
proof shows the persistence of an explicit
short scale correlation structure in the condensate. 
\end{abstract}

\maketitle

Recent experiments on trapped Bose gases in the condensate phase
\cite{CW, Kett} have revived interest in  the rigorous justification
of the Gross-Pitaevskii (GP) theory  from interacting many-body
Hamiltonians. In these experiments, the gas is initially
trapped by a strong magnetic field and  cooled down at very low
temperatures, so that almost all particles condensate in the same
collective mode described by a one-particle wave function,
$\ph(\br)$, $\br\in\bR^3$. Then the traps are instantaneously
removed and the evolution of the condensate is observed.
 To describe this evolution,
the GP theory \cite{G1,P,GP} postulates that
many-body effects can be compressed into a non-linear on-site
interaction.  In units where $\hbar =1$ and the mass of
the bosons $m =1/2$, the GP equation
is given by \be\label{GP}
  i\partial_t\ph_t = -\Delta \ph_t +
 \sigma |\ph_t|^2 \ph_t\; , \ee
with the normalization  $\int |\ph(\br)|^2\rd^3\br=1$. The
coupling constant is $\sigma = 8\pi Na$ where $N$ is the number of
particles and $a$ is the scattering length of the interaction
potential.

A single orbital theory might indicate that the full $N$-body wave
function essentially factorizes. This, however, is usually not the
case: very short scale interactions may introduce strong
pair-correlations that substantially influence the energy of the
system and the dynamics even on larger scales. In particular, the
emergence of the scattering length in $\sigma$ is a  correlation
effect and it is remarkable that this correlation can be
consistently modeled by the same coupling constant along the
whole evolution.

The GP equation thus implicitly postulates a persistent
two-scale
structure of the state. On short scales, the state exhibits a  time
independent pair-correlation,  while on large scales it is given by
the product of  $N$ copies of a time dependent orbital.
It is a fundamental problem in  many-body theory to show that  the GP
postulation of a two-scale structure is a rigorous consequence of the
first principle many-body Schr\"odinger equation.
While many theoretical work addressed this question for
low energy  states, there is no rigorous result from a dynamical point of view.
In this  letter, we shall show
that not only the two-scale structure
is  {\it preserved}, but
it even {\it emerges} dynamically for  a class of initial data.

On the mathematically rigorous level, the GP theory has only been
verified for the ground state of the trapped Bose gas. Lieb,
Seiringer and Yngvason \cite{LSY1} considered a model where the
scattering length $a$ of the interaction varies with $N$, so that
$Na$ remains constant. For repulsive interactions they proved that
the ground state energy per particle in the GP limit ($N\to\infty$,
$a_0=Na=$ fixed) is given by a variational principle, $\min\{
\cE_{\text{GP}}(\ph)\; : \; \| \ph\|=1\}$, where
\begin{equation}\label{eq:GPfun}
\begin{split}
\cE_{\text{GP}}(\ph)= \int \rd^3 \br &\left( |\nabla \ph (\br)|^2 +
V_{\text{ext}} (\br) |\ph (\br )|^2 \right. \\ & \left.+ 4 \pi a_0
|\ph (\br)|^4 \right) \, \end{split} \end{equation} is the GP energy
functional. This result was extended to rotating Bose systems that
model superfluidity in \cite{LSY2}.
In \cite{LS}, Lieb and Seiringer have also rigorously established
that the ground state of the trapped Bose gas exhibits a complete
Bose-Einstein condensation (BEC) by showing that the one particle
density matrix of the ground state converges to the one dimensional
projection $|\ph_{\text{GP}} \rangle \langle \ph_{\text{GP}}|$,
where $\ph_{\text{GP}}$ is  the minimizer of the functional
$\cE_{\text{GP}}$.

The ground state  has been known to have a non-trivial pair
correlation. Mathematically, this was first exploited for the
homogeneous dilute Bose gas, when Dyson \cite{Dy} used a specific
nearest neighbor correlation in his trial state to verify the upper
bound on the ground state energy obtained from Bogoliubov theory
\cite{Landau}. The proof of the lower bound, achieved by Lieb and
Yngvason forty years later \cite{LY}, demonstrates that low energy
states necessarily contain a characteristic short scale pair
correlation. In the GP limit, as shown in \cite{LSY1}, the trapping
potential introduces a second length scale but it leaves the  short
scale structure intact.

The works mentioned above justify the two-scale hypothesis of
the GP theory based on the energy minimizing principle. In this
letter, we outline a mathematical proof (see \cite{ESY1,ESY2} for
details) of the fact that the GP theory also describes the dynamics
of trapped Bose-Einstein condensates. Assuming that at time $t=0$
we have a condensate, we show, under some conditions on the
interaction potential and on the initial state, that the system
still exhibits complete condensation at times $t \neq 0$, and that
the condensate wave function evolves according to the GP equation
(\ref{GP}). Our results cover two types of initial data: type I
states with the physical two-scale structure as in \cite{LSY1,LS}
and type II states with no short scale structure. For type I initial
data, the main difficulty is to demonstrate the persistence of the
two-scale structure even for states that are far from the ground
state. Energy conservation alone does not prevent the destruction of
the short scale structure. The key observation is that higher
moments of the energy are also conserved and they impose stringent
restrictions on the state. In Theorem 2 below we show that a finite
second moment of the energy per particle forces a pair correlation
described by the zero energy scattering function of the interaction
potential. For type II initial data, our results imply that the
short scale structure emerges dynamically. See the discussion after
Theorem 1.

\smallskip

We now define the model precisely. The Hamiltonian for $N$
bosons in three dimensions is given by
\begin{equation}\label{eq:wth}
\wt H_N = \sum_{j=1}^N  (-\Delta_{\br_j}+V_{\text{ext}}(\br_j)\big)
 + \sum_{i<j}^N V_N (\br_i -\br_j)\,.
\end{equation}
The trap potential $V_{\text{ext}} (\br)$ is a positive function
with  $V_{\text{ext}} (\br)\to \infty$
 as $|\br|\to\infty$. The interaction
potential is scaled as $V_N (\br) = N^2 V(N\br)$,  where $V$ is a
positive, spherically symmetric, compactly supported, smooth
potential with scattering length $a_0$. By scaling, the scattering
length of $V_N$ is $a_0/N$. We assume that the strength of the
interaction, measured by the dimensionless quantity $\a=  \int
V(\br)|\br|^{-1}\rd\br +\sup_{\br\in \bR^3} \br^2 V(\br)$,
is sufficiently small.

Let $\psi_N$ be the ground state of (\ref{eq:wth}) and
let $\gamma^{(k)}_N$ be its $k$-particle marginal
density $(1\leq k\leq N$)
with a consistent normalization $\tr \; \gamma^{(k)}_N =1$.
It was proven in \cite{LS} that
\be\label{eq:cond} \gamma^{(k)}_N\to
|\ph_{\text{GP}}\rangle\langle\ph_{\text{GP}}|^{\otimes k}, \qquad
N\to\infty \ee for any fixed $k$. In particular, \eqref{eq:cond} for
$k=1$ means that
 $\psi_N$ exhibits complete
condensation.
After instantaneously removing the trap, the evolution of the system
is generated by the Hamiltonian
\begin{equation}\label{eq:ham1}
 H_N = \sum_{j=1}^N  -\Delta_{\br_j}
 + \sum_{i<j}^N V_N (\br_i -\br_j).
\end{equation}

{\bf Theorem 1.} {\it Let $\psi_{N,t}$ be the solution to the
Schr\"odinger equation $i\partial_t\psi_{N,t} = H_N \psi_{N,t}$ with
the initial condition $\psi_{N,0}=\psi_N$, and let $\gamma_{N,
t}^{(1)}$ be its one particle marginal density. Then, for any fixed
time $t \in \bR$, $\psi_{N,t}$ exhibits complete Bose-Einstein
condensation, that is
\begin{equation}\label{eq:conv}
\gamma^{(1)}_{N,t} \to |\ph_t \rangle \langle \ph_t| \qquad \text{as
} N \to \infty,
\end{equation}
where $\ph_t$ solves the Gross-Pitaevskii equation
\begin{equation}\label{eq:GP}
i \partial_t \ph_t = -\Delta \ph_t+ 8\pi a_0 |\ph_t|^2 \ph_t
\end{equation} with initial data $\ph_{t=0} = \ph_{\text{GP}}$.
The convergence in (\ref{eq:conv})  is in the sense that
$\mbox{Tr}\; K \; (\gamma^{(1)}_{N,t} - |\ph_t \rangle \langle
\ph_t|)\to 0$ for any compact operator $K$ on $L^2(\bR^3)$.
 }

\smallskip

Our proof, in fact, allows for more general initial data.
Let $H^1(\bR^3):=\{ \ph(\br)\; : \; \int |\ph|^2 + |\nabla\ph|^2<\infty\}$.
 The  technical condition we need
to conclude \eqref{eq:conv}, \eqref{eq:GP} is that the initial state
$\psi_{N,0}$ asymptotically factorizes in the sense that there
exists $\ph \in H^1 (\bR^3)$
and for every $k\geq 1$ there exists a
family $\{\xi^{(N-k)}_N \}_{N \geq k}$ with $\xi_N^{(N-k)} \in L^2
(\bR^{3(N-k)})$ and $\| \ph \| = \| \xi^{(N-k)}_N \| =1$ such that
\begin{equation}\label{eq:fact} \lim_{N\to\infty}
\left\| \, \psi_{N,0} - \ph^{\otimes k} \otimes \xi^{(N-k)}_N
\right\| = 0 \end{equation} The function $\ph$ is then the initial
data for \eqref{eq:GP}. As we prove in Appendix C of \cite{ESY2},
the ground state $\psi_N$ of the Hamiltonian (\ref{eq:wth})
satisfies (\ref{eq:fact}). This shows that the pair correlations of
$\psi_N$, which affect the dynamics in a non-trivial way, are not in
conflict with the asymptotic factorization as long as the
convergence in (\ref{eq:fact}) is in a weaker topology than the
energy norm.
The local correlation
structure of $\psi_N$ lives on a length scale $1/N$ and its effect,
measured in $L^2$-sense in \eqref{eq:fact}, is negligible.  However,
the {\it energy} of $\psi_N$ does not converge to the energy of the
factorized state, and, similarly, the convergence in \eqref{eq:cond}
does {\it not} hold in the $H^1$ space or in energy sense.

A typical type II initial data is given by a product wave
function $\psi_{N,0} = \ph^{\otimes N}$ with  $\ph \in H^1 (\bR^3)$.
For $\psi_{N,0}$ the condition \eqref{eq:fact} holds trivially. Our
result thus shows that a product initial state also evolves
according to the GP equation \eqref{eq:GP} (with initial data
$\ph_{t=0} = \ph$) despite that its total energy is  larger than
$\cE_{\text{GP}}(\ph)$. In fact, it is easy to see that the
energy per particle in the factorized state, $\langle \ph^{\otimes
N}, (\wt H_N / N) \ph^{\otimes N}\rangle$, is asymptotically given
by a functional similar to \eqref{eq:GPfun} but with $8\pi a_0$
replaced with its Born approximation, $b_0=\int V(\br)\rd^3 \br$.
Note that $b_0>8\pi a_0$. This indicates that the excess energy
originating from the emergence of the scattering length is dispersed
into modes that do not influence the evolution of the condensate.
Hence, for product initial data, the scattering length emerges
dynamically and the energy functional of the single particle
orbitals does not predict the evolution equation.

\smallskip

{\it Outline of the proof.} For $k=1,\dots,N$, let
$\gamma^{(k)}_{N,t}$ denote the $k$-particle marginal density of
$\psi_{N, t}$ with a consistent normalization $\tr \;
\gamma^{(k)}_{N,t} =1$. The time evolution is governed by a
hierarchy of $N$ coupled equations, commonly known as the BBGKY
hierarchy
\begin{equation}\label{eq:BBGKY}
\begin{split}
i\partial_t &\gamma^{(k)}_{N,t} = \; \sum_{j=1}^k [-\Delta_{\br_j} ,
\gamma^{(k)}_{N,t} ] + \sum_{i<j}^k [V_N (\br_i -\br_j) ,
\gamma^{(k)}_{N,t}] \\ &+ (N-k) \sum_{j=1}^k \tr_{k+1} \; [ V_N
(\br_j - \br_{k+1}) , \gamma^{(k+1)}_{N,t}]
\end{split}
\end{equation}
where $\tr_{k+1}$ denotes the partial trace over $\br_{k+1}$.
By the Banach-Alaouglu theorem, $\gamma^{(k)}_{N,t}$ has at least
one weak* limit point, $\gamma_{\infty,t}^{(k)}$, in the space of
trace class operators. Since formally $N V_N (\br) = N^3 V (N\br)
\to b_0 \delta (\br)$, one may naively expect from
(\ref{eq:BBGKY}) the time evolution of $\gamma_{\infty,t}^{(k)}$ to
be described by the infinite hierarchy
\begin{equation}\label{eq:infhier}
\begin{split}
i\partial_t  \gamma^{(k)}_{\infty,t} = &\sum_{j=1}^k [-\Delta_{\br_j} ,
\gamma^{(k)}_{\infty,t} ] \\& + \sigma \sum_{j=1}^k \tr_{k+1}
[\delta (\br_j -\br_{k+1}), \gamma^{(k+1)}_{\infty,t}]
\end{split}
\end{equation} with $\sigma = b_0$. However, the coupling
constant in the correct equations for $\gamma^{(k)}_{\infty,t}$ is
$\sigma = 8 \pi a_0$. The reason is that $\gamma^{(k+1)}_{N,t}$
develops a short scale structure which lives on the same length
scale $\ell \simeq 1/N$ as the potential $V_N$. Since
$\gamma^{(k+1)}_{N,t}$ is not constant on this length scale, one
cannot apply the formal limit $N V_N \to b_0 \delta $ in
(\ref{eq:BBGKY}).

In the following Theorem 2 we show that for $\br_j$ close to $\br_{k+1}$, the
short scale structure of $\gamma^{(k+1)}_{N,t}$ is described by the
function $f_N (\br_j -\br_{k+1})$, where $f_N$ is the solution to
the zero energy scattering equation
\begin{equation}\label{eq:scatt}
\left(-\Delta + \frac{1}{2} V_N (\br) \right) f_N (\br) = 0
\end{equation} with boundary condition $f_N (\br) \to 1$ as $|\br| \to
\infty$. By scaling, $f_N(\br) = f(N\br)$, where $f$ is a
solution of $\left(-\Delta + \frac{1}{2} V \right) f = 0$
with the same boundary condition.
Therefore the correct value of
$\sigma$ in (\ref{eq:infhier}) is given by
\[ \sigma = \int \rd^3 \br \, N V_N (\br) f_N (\br)
= \int \rd^3 \br \, V(\br) f(\br) = 8 \pi a_0 \, .\] Note that the
short scale structure disappears from $\gamma^{(k)}_{\infty,t}$
after taking the weak limit, but it still affects the limiting
macroscopic dynamics.

The infinite hierarchy (\ref{eq:infhier}) has a factorized solution:
the family $\gamma^{(k)}_{\infty,t} = |\ph_t \rangle \langle
\ph_t|^{\otimes k}$, $k \geq 1$, is a solution to (\ref{eq:infhier})
with $\sigma = 8 \pi a_0$ if and only if $\ph_t$ is a solution to
the GP equation (\ref{eq:GP}). Therefore, to identify the limiting
density, it suffices to show that (i) any limit point $\{
\gamma^{(k)}_{\infty,t}\}_{k \geq 1}$ of the family of densities $\{
\gamma^{(k)}_{N,t} \}_{k=1}^N$ is a solution of the infinite
hierarchy (\ref{eq:infhier}) with $\sigma = 8 \pi a_0$, and (ii) the
solution to the infinite hierarchy is unique.  The proofs of (i),
(ii) rely on estimates on the energy distribution of the initial
wave function and on the factorization property (\ref{eq:fact}).

\smallskip

{\it (i) Convergence to the infinite hierarchy.} The following key
theorem identifies the short scale structure assuming a bound on the
second moment of the energy.  This is the main ingredient in the
proof of (\ref{eq:infhier})

\smallskip

{\bf Theorem 2.} {\it Suppose $\a$ is small enough. Then there
exists a constant $C >0$ such that
\begin{equation}\label{eq:ee}
\langle \psi , H_N^2 \psi \rangle \geq C N^2 \int \rd \bRR \, \left|
\, \nabla_{\br_i} \nabla_{\br_j} \frac{\psi (\bRR)}{f_N (\br_i
-\br_j)} \right|^2
\end{equation}
for all $1\leq i< j\leq N$ and all $\psi \in L^2 (\bR^{3N}, \rd
\bRR)$ symmetric with respect to permutations. Here $\bRR = (\br_1,
\dots, \br_N) \in \bR^{3N}$.}

\smallskip

{F}rom this theorem, we obtain the a-priori bound
\begin{equation}\label{eq:apri} \int \rd \bRR
\left|\, \nabla_{\br_i} \nabla_{\br_j} \frac{\psi_{N,t} (\bRR)}{f_N (\br_i
-\br_j)} \right|^2 \leq C, \quad \forall \; i \neq j\end{equation}
uniformly in $N$ and $t$ provided the initial wave function
$\psi_{N,0}$ at time $t =0$ satisfies $\langle \psi_{N,0} , H_N^2
\psi_{N,0} \rangle \leq C N^2$. Note that $\int \rd\br_i\rd\br_j
|\nabla_{\br_i}\nabla_{\br_j}
 f_N(\br_i-\br_j)|^2 \sim N$ thus (\ref{eq:apri})
identifies the short scale structure of $\psi_{N,t}$ with a
precision $1/N$. Furthermore, since the left side of (12) is a constant
of motion, (\ref{eq:apri})  shows that the
separation between the singular short scale structure and the
regular part of $\psi_{N,t}$ is preserved by the time evolution.

\smallskip

{\it Outline of the proof of Theorem 2:} we define \( h_m =
-\Delta_{\br_m} + \frac{1}{2} \sum_{n \neq m} V_N (\br_n -\br_m) \)
for $m=1,\dots,N$. Then $H_N = \sum_{m=1}^N h_m$ and, using the
permutation symmetry of $\psi$ and  $V \geq 0$, we obtain, for
arbitrary $i \neq j$,
\begin{equation*}
\begin{split}
&\frac{\langle \psi, H_N^2 \psi \rangle}{N(N-1)} \geq  \langle \psi
, h_i h_j \psi \rangle
\\ &\geq  \Big\langle \psi, (-\Delta_{\br_i} + \frac{V_N (\br_i
-\br_j)}{2}) 
(-\Delta_{\br_j} + \frac{V_N (\br_j -\br_i)}{2}) \psi \Big\rangle .
\end{split}
\end{equation*}
{F}rom the definition (\ref{eq:scatt}) of $f_N (\br)$, we find
\begin{equation*} (-\Delta_{\br_i} + \frac{1}{2} V_N (\br_i -\br_j))
 \psi = f_N
(\br_i-\br_j) L_i \frac{\psi}{f_N (\br_i -\br_j)} \end{equation*}
with $L_i = -\Delta_{\br_i} + 2\nabla_{\br_i} (\log  f_N (\br_i
-\br_j)) \nabla_{\br_i}$ and an analogous identity for $L_j$.
 Using integration by parts, and with $
\phi_{ij} (\bRR) = \psi (\bRR) / f_N (\br_i -\br_j)$,
 \begin{equation*}
\begin{split} &\frac{\langle \psi, H_N^2 \psi \rangle}{N(N-1)}
\geq \int \rd \bRR \; f_N^2 (\br_i -\br_j) | \, \nabla_{\br_i}
\nabla_{\br_j} \phi_{ij} (\bRR)|^2 \\&+ \int \rd \bRR \; \left(f_N^2
\nabla^2 \log f_N \right) (\br_i -\br_j) \, \nabla_{\br_i}
\overline{\phi}_{ij} (\bRR) \nabla_{\br_j} \phi_{ij} (\bRR).
\end{split}
\end{equation*}
Eq. (\ref{eq:ee}) now follows because $|\nabla^2 \log f_N (\br_i
-\br_j)| \leq C \a \; |\br_i - \br_j|^{-2}$ and therefore the second term
on the right hand side of the last equation can be controlled by the
first one (for $\a$ small enough) using the operator inequality
$|\br|^{-2} \leq - C \Delta_\br$ (Hardy inequality).

\smallskip

{\it (ii) Uniqueness of the infinite hierarchy.} The first step
is to prove a-priori bounds
in a certain Sobolev norm.

{\bf Theorem 3.} {\it Let $\gamma^{(k)}_{\infty,t}$ be any
weak limit point of $\gamma^{(k)}_{N,t}$, then
the following estimate holds uniformly in time
\begin{equation}\label{eq:aprik}
\tri\gamma^{(k)}_{\infty,t}\tri_k:=
 \tr \; (1-\Delta_{\br_1}) \dots (1-\Delta_{\br_k})
\gamma^{(k)}_{\infty,t} \leq C^k.
 \end{equation}}
\smallskip
{\it Idea of the proof.}
By conservation of $H_N^k$ along the evolution,
$\tr H_N^k
\gamma_{N,t}^{(k)} = \tr H_N^k \gamma_{N,0}^{(k)}$. A
control on  $\tr H_N^k \gamma_{N,0}^{(k)}$ can be obtained
through \eqref{eq:fact}.
 The difficulty   is that
the norm $\tri\cdot\tri_k$ cannot be directly controlled by $\tr
H_N^k(\cdot)$
 and in fact  $\tri
\gamma^{(k)}_{N,t}\tri_k\to\infty$ as $N\to \infty$
because of the singular short scale structure.
It is only after taking the weak limit $N\to\infty$
that the short scale structure
disappears and (\ref{eq:aprik}) can be proven.

For illustration, consider the case $k=2$ discussed in Theorem 2.
The estimate  \eqref{eq:apri} implies that
$\psi_{N,t} (\bRR) \sim {f_N (\br_i -\br_j)} \phi_t (\bRR)$ where
$\phi_t$ is a smooth function in the variable $ \br_i -\br_j$.
Together with the fact that $\int \rd\br_i\rd\br_j
|\nabla_{\br_i}\nabla_{\br_j}
f_N(\br_i-\br_j)|^2 \sim N$, we have  that
$  \int \rd \bRR \,
\left | \, \nabla_{\br_i} \nabla_{\br_j}  \psi_{N}(\bRR) \right |^2
\to \infty $ as $N\to \infty$.
However, since $f_N\to 1$ weakly in $L^2$-sense (but not in
energy sense), a Sobolev estimate
on the limit of the density matrix of $\psi_{N,t}$ can be deduced from
\eqref{eq:apri}.

\smallskip

The uniqueness holds in the Sobolev norm $\tri \cdot \tri_k$:

\smallskip

{\bf Theorem 4.} {\it Given a family of  densities
$\Gamma = \{ \gamma^{(k)} \}_{k \geq 1}$ such that
$\tri  \gamma^{(k)}\tri_k\leq C^k$,
 there exists at most one solution
$\Gamma_t = \{ \gamma^{(k)}_t \}_{k \geq 1}$ to (\ref{eq:infhier})
with $\Gamma_{t=0} = \Gamma$ and such that $\tri \gamma_t^{(k)}\tri_k\leq C^k$ holds
 uniformly in $t$. }

\smallskip

{\it Outline of the proof.} Iterating the integral form of
(\ref{eq:infhier}),
we obtain a Dyson series
\begin{equation}\label{eq:dyson}
\gamma^{(k)}_t = \cU^{(k)} (t) \gamma^{(k)} + \sum_{m=1}^{n-1}
\omega^{(k)}_{m,t},
+ \eta_{n,t}^{(k)},
\end{equation}
where $\cU^{(k)}_t\gamma^{(k)} = e^{it\sum_{j=1}^k \Delta_j}
\gamma^{(k)} e^{-it\sum_{j=1}^k \Delta_j}$, and
\begin{equation}
\begin{split}\label{om}
\omega^{(k)}_{m,t} = \; &\int_0^t \rd s_1 \dots \int_0^{s_{m-1}} \rd
s_m \, \cU^{(k)}_{t-s_1} B^{(k)} \\ & \dots \times B^{(k+m-1)}
\cU^{(k+m)}_{s_m} \gamma^{(k+m)},
\end{split}
\end{equation}
with the collision operator given by
\[ B^{(k)} \gamma^{(k+1)} = -i \sigma
\sum_{j=1}^k \tr_{k+1} \; [ \delta (\br_j - \br_{k+1}) , \gamma^{(k+1)}
] \,. \] The error term $\eta^{(k)}_{n,t}$  has
the same form as $\omega^{(k)}_{n,t}$
with $\cU^{(k+n)}_{s_n} \gamma^{(k+n)}$  replaced by the full evolution
$\gamma^{(k+n)}_{s_n}$.

\begin{figure}
\begin{center}
\epsfig{file=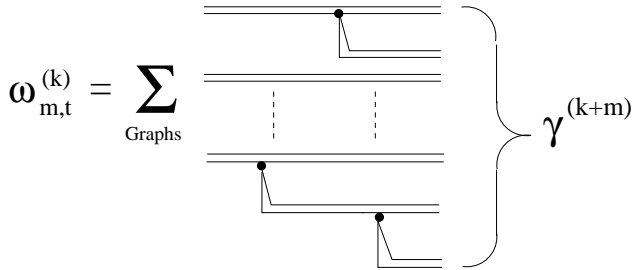,scale=.52}
\end{center}
\caption{Diagrammatic expansion for $\omega_{m,t}^{(k)}$ as sum of
Feynman graphs with $m$ vertices, $2(k+m)$ leaves and
 $2k$ roots} \label{fig:feynman}
\end{figure}

To prove the convergence of  the expansion \eqref{eq:dyson} as
$n\to\infty$,  we expand each term into a sum of contributions
associated with certain Feynman graphs. A typical graph $\Lambda$
contributing to $\omega_{m,t}^{(k)}$ is drawn in Fig.
\ref{fig:feynman}. It has $m$ four-valent vertices and $2k+3m$
lines. The external lines on the left (called {\it roots})
correspond to the $2k$ momenta variables of the operator kernel of
$\omega_{m,t}^{(k)}$. The $2(k+m)$ external lines on the right
(called {\it leaves}) represent the kernel of $\gamma^{(k+m)}$. The
graphical structure of $\Lambda$ encodes the collision history given
in \eqref{om}. Every line $e$ of $\Lambda$ carries a regularized
free propagator, $(\a_e - \bp_e^2 \pm i/t)^{-1}$ with a momentum
variable $\bp_e\in \bR^3$ and a frequency $\a_e\in \bR$. At each
vertex, there is a $\bp$- and a $\a$-delta function due to momentum
and energy conservation. The kernel of $\omega_{m,t}^{(k)}$ (whose
 variables correspond to the $2k$ momenta of the roots) is
computed by performing $3m$ momentum integrals and $2k+3m$ frequency
integrals in each $\Lambda$.

Because of the singularity of the interaction at $\br=0$, each graph
is potentially ultraviolet divergent. Power counting suggests,
however, that the integrals are finite.
Suppose we cutoff all momentum integrals at $|\bp| \simeq \beta\gg
1$, and all frequency integrals at $|\a| \simeq \beta^2$; then the
integration volume scales as $\beta^{3(3m) +
2(2k+3m)}=\beta^{4k+15m}$. The $m$ momentum and frequency delta
functions scale as $\beta^{-5m}$ and the $2k+3m$ propagators scale
as $\beta^{-2(2k+3m)}$. The a-priori estimate (\ref{eq:aprik}) shows
that $\gamma^{(k+m)} \lesssim \beta^{-5(k+m)}$. Since $4k+15m < 5m +
2 (2k+3m) + 5 (k+m)$, the integrals should be convergent in the
ultraviolet regime $\beta \gg 1$. To make this  argument rigorous,
we use an integration scheme, dictated by the structure of the
graph. We start by integrating the momenta and frequency of the
leaves; these integrals are convergent because of the a-priori
estimates (\ref{eq:aprik}) and they also provide a momentum decay on
the lines adjacent to the leaves. We then iterate this procedure,
integrating all momenta and frequencies by moving from the right to
the left of the graph and transferring a suitable momentum decay.

\smallskip

{\it Conclusion.} We have proven that  Bose-Einstein condensates
evolve according to the Gross-Pitaevskii equation. This provides a
mathematical description of recent experiments on the evolution of
initially trapped condensates. On the theoretical level, our result
for factorized initial wave functions is more surprising. The
emergence of the scattering length in the GP equation is a
consequence of the short scale correlations. We show, however, that
the GP equation is correct even if the initial state is uncorrelated
(product). Our result thus suggests that the many body dynamics
builds up correlations on the length scale $\ell \simeq 1/N$ in a
very short time. Since the emergence of correlations reduces the
local energy, one may ask what happens with the excess energy.
Although we are not able to give a mathematically rigorous answer,
we believe that the excess energy is transferred to incoherent
excitations living on intermediate length scales $\ell$, with
$N^{-1}\ll \ell\ll 1$. Since the macroscopic dynamics described by
the GP equation is affected only by the structure of the wave
function on length scales of $O(1)$ and  $O(1/N)$, these mesoscopic
excitations have no influence on the evolution of the condensate.

\smallskip

{\it Acknowledgements.} The work of H.-T. Yau was partially
supported by NSF grant DMS-0602038.

\end{document}